\documentclass[letterpaper, 10 pt, conference]{ieeeconf}  

\IEEEoverridecommandlockouts                        

\usepackage{graphics} 
\usepackage{epsfig} 
\usepackage{mathptmx} 
\usepackage{times} 
\usepackage{amsmath} 
\usepackage{amssymb}  
\usepackage{graphicx, tabularx, bm }
\usepackage{subfig}
\usepackage{floatrow}

\title{\LARGE \bf
Global Perturbation of Initial Geometry in a Biomechanical Model of Cortical Morphogenesis 
}

\author{A.Bohi$^{1}$, X.Wang$^{2}$, M.Harrach$^{3}$, M.Dinomais$^{3}$, F.Rousseau$^{2}$, J.Lef\`{e}vre$^{1}$
\thanks{$^{\ast}$ The research leading to these results has been supported by the Fondation pour la Recherche M\'{e}dicale (grant DIC20161236453).}
\thanks{$^{1}$ Institut de Neurosciences de la Timone, Aix-Marseille Universit\'e, CNRS UMR 7289. Marseille, France. 
        {\tt\small \{amine.bohi, julien.lefevre\}@univ-amu.fr}}
\thanks{$^{2}$ IMT Atlantique, LaTIM U1101 INSERM, UBL, Brest,France.{\tt\small xiaoyu.wang@imt-atlantique.fr, francois.rousseau@telecom-bretagne.eu, }}
\thanks{$^{3}$ Laboratoire Angevin de Recherche en Ing\'{e}nierie des Syst\`{e}mes (LARIS), Angers, France.                {\tt\small mariam.harrach@hotmail.com, mickael.dinomais@gmail.com}}
}

\begin{document}

\maketitle
\thispagestyle{empty}
\pagestyle{empty}

\begin{abstract}

Cortical folding pattern is a main characteristic of the geometry of the human brain which is formed by gyri (ridges) and sulci (grooves).
Several biological hypotheses have suggested different mechanisms that attempt to explain the development of cortical folding and its abnormal evolutions. Based on these hypotheses, biomechanical models of cortical folding have been proposed. In this work, we compare biomechanical simulations for several initial conditions by using an adaptive spherical parameterization approach. Our approach allows us to study and explore one of the most potential sources of reproducible cortical folding pattern: the specification of initial geometry of the brain.

\end{abstract}

\section{INTRODUCTION}
\label{sec:intro}

The cerebral cortex, also called the grey matter, is geometrically characterized by its folds, i.e. spatial alternation of gyri and sulci. In the literature, there have been two contrasting biological theories that attempt to explain the formation of cortical folding. Van Essen \cite{van1997tension} proposes that mechanical tension along axons in white matter is a major driving force for many aspects of the central nervous system morphogenesis. Richman et al. \cite{richman1975mechanical} suggests that a mechanical buckling induced by differential expansion - with a faster growing grey matter and a slower growing white matter - leads to cortical folding.

Based on these biological theories, biomechanical models of cortical folding have been proposed. The most recent one has been introduced based on the axonal tension-based hypothesis \cite{budday2014mechanical} with limited validations, restricted to qualitative comparisons with real data. Another recent model \cite{tallinen2016growth,tallinen2014gyrification} has implemented the mechanism suggested by the hypothesis of differential growth. The strength of the last model lies in the use of magnetic resonance images (MRI) of the developing smooth fetal brain, which allows us to specify a realistic initial geometry to numerical simulations. 

When studying a numerical model the sensitivity to initial conditions is crucial. For instance in \cite{zhang2016mechanism} local perturbations in parameters revealed different folding patterns. However, the impact of the initial brain geometry did not give rise to a particular study. It is already known that the degree of cortical folding is dependent on the overall size of the brain \cite{toro2008brain}, and that this strong relationship can be modified by brain developmental pathologies such as Microcephaly \cite{germanaud2014simplified}. Another natural question is to determine whether variations in the global shape of the brain, independently of scale, have an impact on the surface morphology generated by a biomechanical model. 

In this paper, following a line of research started in \cite{tallinen2016growth,tallinen2014gyrification,lefevre2015spherical}, we present a framework that allows us, first, to analyse the evolution of the global shape of the brain across biomechanical simulations, and then to study the impact of the initial brain geometry on surface morphology using an adaptive spherical mapping approach to measure similarities between several simulated cortical surfaces generated by biomechanical model.

\section{METHODS}
\label{sec:format}

\subsection{Biomechanical model of brain convolution development}
\label{ssec:biomeca}
Biomechanical simulation strives to reproduce physical behaviour
of a living human brain represented by a physical model. If the latter proposes a geometrical representation of this organ, it also includes a set of mechanical properties describing its physical behaviour.

In \cite{tallinen2014gyrification}, the authors show that the gyrification in the human brain can occur as a non-linear consequence of a mechanical instability driven by compressive stress due to the differential tangential expansion of the cortical layer constrained by the subcortical one. The simulation of cortical folding is based on a finite element model of the human brain with a 3D tetrahedral mesh extracted from a T2-weighted motion corrected MRI of a fetal brain at 22 GW. The cortical surface is first triangulated and the brain volume is meshed uniformly with approximately $2*10^{7}$ tetrahedrons. The brain is modelled as a compressible neo-Hookean hyperelastic material with a volumetric strain energy density. We will not go into the details here, and the interested reader can refer to \cite{tallinen2016growth,tallinen2014gyrification}

Each local element (Tetrahedrons) grows according to the growth tensor $G$ that describes the differential tangential expansion perpendicular to the surface normal and is defined by

\begin{equation}
  G = gI+(1-g)\hat{n}_s\otimes{\hat{n}_s},
\end{equation}

where $n_s$ is the surface normal of the undeformed reference state of the simulation and $g$ is the tangential expansion ratio of the cortical layer compared to the subcortical one. In \cite{tallinen2014gyrification}, this measure is defined by

\begin{equation}
  g(y) = 1 + \frac{\alpha}{1+\exp(10(\frac{y}{H}-1))},
\end{equation}

with $g=1$ in the subcortical layer and $g=1+\alpha$ in the cortical one, $\alpha$ represents the magnitude of the expansion, $y$ is the distance from the top surface and $H$ denotes the undeformed cortical thickness. Since this tangential growth is constrained and governed by material behaviour and expansion, it induces mechanical stresses and consequently generates elastic forces in the affected areas (tetrahedrons and their neighboring), causing the cerebral cortex to deform. The reader can refer again to the articles \cite{tallinen2016growth,tallinen2014gyrification}.

Finally, a deformation volumetric energy of the biomechanical model is minimized using a powerful explicit solver (Newton-Raphson method).

In addition to studying this type of biomechanical model based on the hypothesis of differential growth, our main objective is to use it so that we can disturb and distort the input data (brains tetrahedral meshes) of the biomechanical system and obtain as output data several smooth and folded cortical surfaces. A detailed explanation of this manipulation is presented in \ref{ssec:comparison}.

\subsection{Spherical parameterization using Laplace-Beltrami eigenfunctions}
\label{ssec:mapping}
Spherical parameterization of 2D mesh data is a robust technique for many computer graphics applications, in particular for texture mapping, remeshing and morphing \cite{sheffer2007mesh}, which consists on associating a given closed surface with a spherical domain. It is becoming more and more popular in the neuroimaging community, in accordance with the specific advantages for e.g. inter-subject mapping and visualization \cite{fischl1999cortical}. However the underlying numerical methods can be time consuming which is a clear limitation when it comes to compare a large number of surfaces generated by numerical simulations.

In order to study the impact of initial geometry on gyration, we adapted the fast and simple approach introduced in \cite{lefevre2015spherical} to simulated fetal brain surfaces generated by the biomechanical model described in \ref{ssec:biomeca}. 

\begin{figure}[htp]
\centering
\includegraphics[width=1\textwidth]{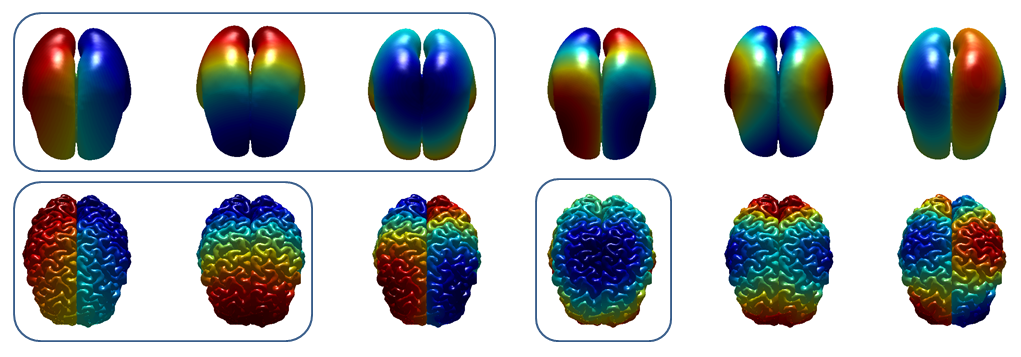}
\vspace{-0.5cm}
\caption{Six first eigenfunctions for a smooth fetal brain (first row) and a simulated cortex (second row). In each row three first non-trivial eigenfunctions with only 2 nodal domains are framed. Colormap goes from blue (-) to red (+).}
\label{fig:eigen}       
\end{figure}

In \cite{lefevre2015spherical}, the authors introduced a robust approach for directly parameterizing a certain class of closed surfaces without holes (genus-zero) onto a spherical domain. They define a mapping by considering only the three first non-trivial eigenfunctions of the Laplace-Beltrami operator and which have exactly two nodal domains. Let us recall that, an eigenfunction of a linear operator $A$, defined on a functional space, is an eigenvector of the operator. This is a non-zero function satisfying: $Au_{i}=\lambda_{i}u_{i}$, for some scalar eigenvalue $\lambda_{i}$ associated to $u_{i}$. We also recall that, by definition, the nodal domains of an eigenfunction $u$ are the connected components of the complement of the nodal set $\Omega_{0} \backslash N(u)$, where $N(u)=\left\{x\in\Omega_{0} | u(x)=0 \right\}$. As an illustration, it corresponds to the number of blue and red spots on the textures of Fig. \ref{fig:eigen}.

Based on preliminary mathematical and empirical results on spectral theory and Laplace-Beltrami eigenfunctions \cite{zelditch2009local,lefevre2015spherical}, the authors advance the following conjecture allowing them to define a natural spherical mapping: \\

\textbf{Main conjecture}  \textit{Let us fix a closed surface $S \subset \mathbb{R}^3$ and consider $u_1$, $u_2$ and $u_3$ the three first eigenfunctions of the Laplace-Beltrami operator associated to the three first nonvanishing eigenvalues. We assume they have only two nodal domains. Then the mapping} 
\begin{equation}
\label{eq:mapping}
\begin{split}
    u: &S \rightarrow \mathbb{S}^2 \\
    &x \rightarrow \frac{1}{r(x)} (u_1(x),u_2(x),u_3(x))
\end{split}
\end{equation}
with $r(x)=(u_1(x)^2+u_2(x)^2+u_3(x)^2)^{-1/2}$
\textit{is well defined and it is a $C^\infty$ diffeomorphism.} \\

\textit{Remark.} In \cite{lefevre2015spherical}, the restriction on the topology of the eigenfunctions guarantees a spherical mapping with only the three first eigenfunctions. However, we can find complex examples of closed surfaces where some of the three first eigenfunctions have more than two nodal domains. In our study, for a folded cortical surface, the third eigenfunction has four nodal domains, whereas the smooth cortical surfaces are very suitable because they satisfy the conditions proposed in \cite{lefevre2015spherical}, with a two nodal domains (see Fig. \ref{fig:eigen}).

In order to overcome the limitation mentioned in the previous remark, we generalize the previous conjecture assuming that, for a genus-zero surface, we can always find three eigenfunctions associated to larger eigenvalues in the spectrum with only two nodal domains, which allows us to provide a mapping similar to \ref{eq:mapping}.

\subsection{Framework to compare different simulations}
\label{ssec:comparison}
\begin{figure*}[thpb]
\centering
  \includegraphics[height=6.13cm,width=0.9\textwidth]{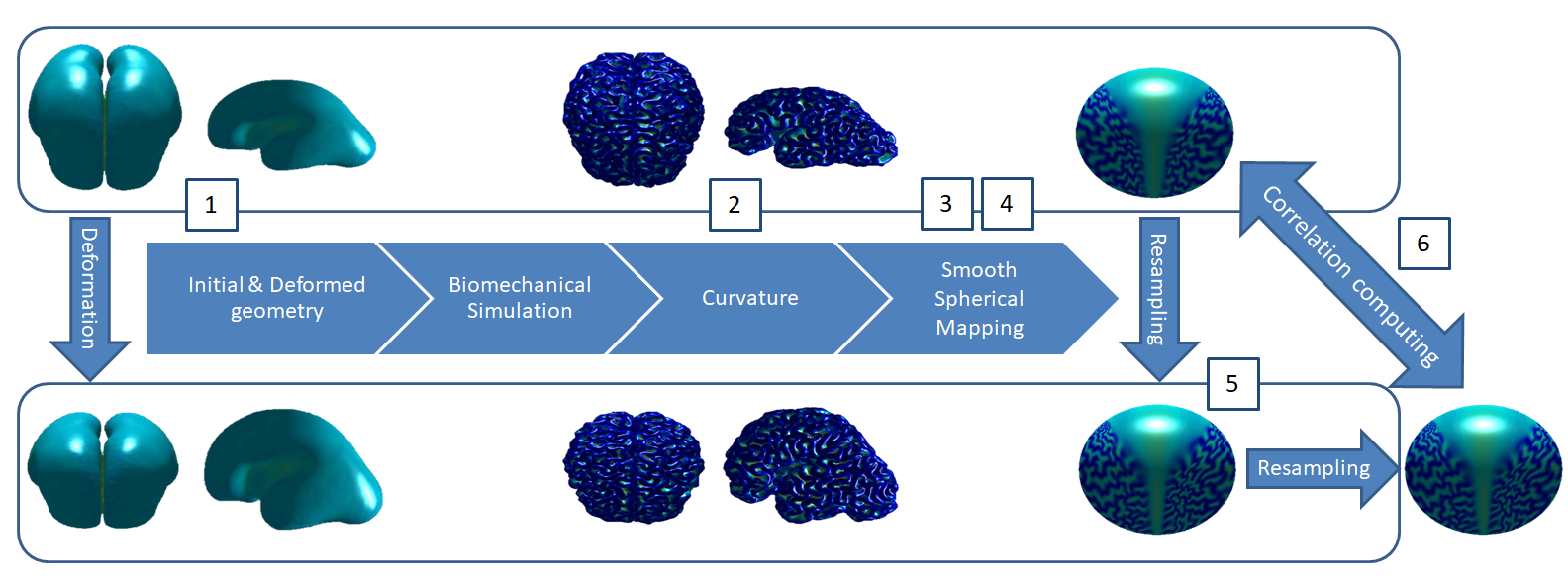}
\vspace{-0.2cm}
\caption{The pipeline of the proposed framework of comparing two brains surfaces. Colormap goes from blue (-) to red (+).}
\label{fig:schema}       
\end{figure*}

In order to study the impact of the initial geometry of the human brain on surface morphology during the cortical development process, we compare different surfaces of fetal brains generated by the biomechanical model based on the finite element model of differential cortical and subcortical growth introduced in \cite{tallinen2014gyrification}, in which initial shapes were derived from in early utero MRI data collected from human fetuses.

In this study, we choose the three most important steps of the biomechanical model, namely: step 500 where the brain is still smooth, step 9000 where the primary folds begin to appear and step 22000 which knows the birth of secondary and tertiary folds.

From these facts, a pipeline for studying the impact of initial brain geometry on surface morphology (see Fig. \ref{fig:schema}) is the following:

\begin{enumerate}
\item Starting from a reference brain $B_{ref}$ (3D brain tetrahedral mesh presented in \ref{ssec:biomeca}), we generate $B_{a,b}$ a deformed anatomy by applying to $B_{ref}$ an affine transformation matrix defined by
\begin{equation} \label{eq:7}
    M_{a,b}=\begin{bmatrix}a & 0 & 0 \\0 & b & 0\\0 & 0 & c\end{bmatrix}
\end{equation}
where $c=\frac{1}{ab}$ is chosen so that the initial perturbed cerebral volume remains constant. Using the model described in \ref{ssec:biomeca}, we generate cortical surfaces $S_{ref}(t)$ and $S_{a,b}(t)$, respectively, from the initial brain geometries $B_{ref}$ and $B_{a,b}$. Note that $S_{ref}(t)=S_{1,1}(t)$.
	
\item We calculate curvatures of $S_{ref}(t)$ and $S_{a,b}(t)$ using an approximation of mean curvature and a computation of the average angle between the normal to the surface of a vertex and the edges formed by the vertex and its neighbours.
	
\item In order to successfully reduce noise from the future spherical maps, a smoothing technique based on heat equations is applied to the curvatures of $S_{ref}(t)$ and $S_{a,b}(t)$.
	
\item Using the adaptive spherical parameterization, we consider the six first eigenfunctions of the Laplace-Beltrami Operator of the closed surfaces $S_{ref}(t)$ and $S_{a,b}(t)$, then we chose three non-trivial eigenfunctions with only two nodal domains and we sort them according to their correlation with x, y and z axes. Finally, we use the selected eigenfunctions to transform the simulated cortical surfaces into spherical maps.
	
\item To complete this process, we resample the texture of the curvature of one of the spherical maps on the other.
	
\item Finally, we use the Pearson correlation coefficient to measure the similarity between the curvature of the surface $S_{ref}(t)$ and the resampled one of the surface $S_{a,b}(t)$ 
\end{enumerate}
Note that, at the end of the step 1, we apply $M_{a,b}^{-1}$ to each surface $S_{a,b}(t)$ generated by the model, in order to have more consistency between the global shapes before the spherical parameterization.

\section{RESULTS}
\label{sec:results}

\subsection{Evolution of the global shape of the brain across simulations}

For each simulation corresponding to the perturbation of the initial shape and for each time step $t$, we estimated an affine matrix $M(t)$ that optimizes the fit between the surface $S_{a,b}(t)$ and the surface in the reference simulation $S_{ref}(t)$. Given that $M(0)$ equals the deformation matrix of equation \ref{eq:7}, the Frobenius norm of $M(t)-M(0)$ quantifies the way the global shape of the brain evolves along a simulation. In absolute value, this norm did not exceed $0.04$ which corresponds to a relative value lower than $2.5\%$ over all the simulations and all the time steps.\\
This result allows us to conclude that 1) applying $M_{a,b}^{-1}$ in \ref{ssec:comparison} is legitimate and that 2) the biomechanical model preserves the global shape of the brain, in spite of the appearance of cortical folding pattern.

\subsection{Impact of the initial brain geometry on surface morphology}
\label{ssec:initgeo}

\begin{figure*}[htp]
\centering
\begin{minipage}{1\textwidth}
\centering
\hspace*{\stretch{1}}%
\subfloat{\includegraphics[width=5.7cm,height=6.4cm]{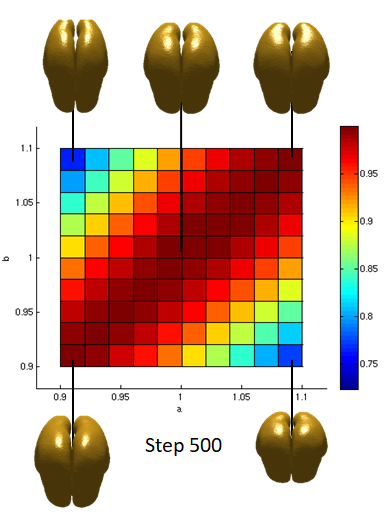}}%
\hspace{\stretch{2}}%
\subfloat{\includegraphics[width=5.7cm,height=6.4cm]{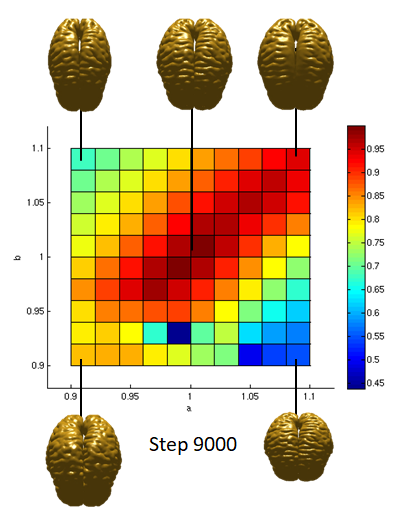}}%
\hspace{\stretch{1}}
\subfloat{\includegraphics[width=5.7cm,height=6.4cm]{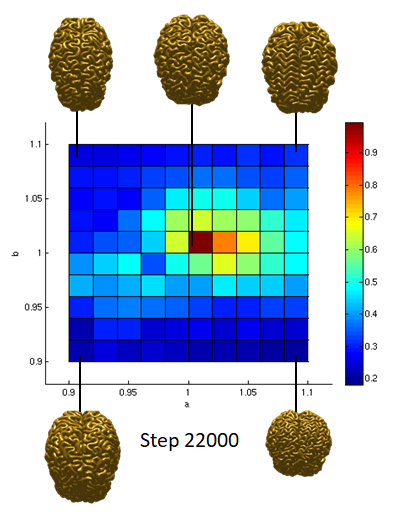}}%
\caption{Correlation values for different scale factors at step 500, 9000 and 22000}
\label{fig:corrab}
\end{minipage}%
\end{figure*}



In order to have small variations in the shape of the brain at the beginning of the tangential expansion process, we apply the transformation matrix $M_{a,b}$ with $(a,b)\in[0.9:0.02:1.1]$.\\
In Fig. \ref{fig:corrab}, we compute the correlation coefficients between cortical surfaces $S_{ref}$ and $S_{a,b}$ for different factors $(a,b)\in[0.9:0.02:1.1]$ and at simulation steps 500, 9000 and 22000. At step 500, we have strong correlations that vary between 0.75 and 1, for almost all scale factors $(a,b)$. This can be explained by the fact that at the beginning of the biomechanical simulation, cortical surfaces are still smooth, which does not influence the similarity between the two surfaces $S_{ref}$ and $S_{a,b}$. At step 9000, correlation values vary between 0.45 and 1, with a best similarity between $S_{ref}$ and $S_{a,b}$ for $(a,b)$ around 1 (identity matrix). This can be interpreted by the fact that at step 9000, the brain folds begin to appear and are also different from a simulation to another. Finally, at step 22000, when the folds become more apparent, we can notice that correlation values fall except for values of $(a, b)$ between 0.98 and 1.02. 

\begin{figure}[htp]
\centering
  \includegraphics[width=1\textwidth]{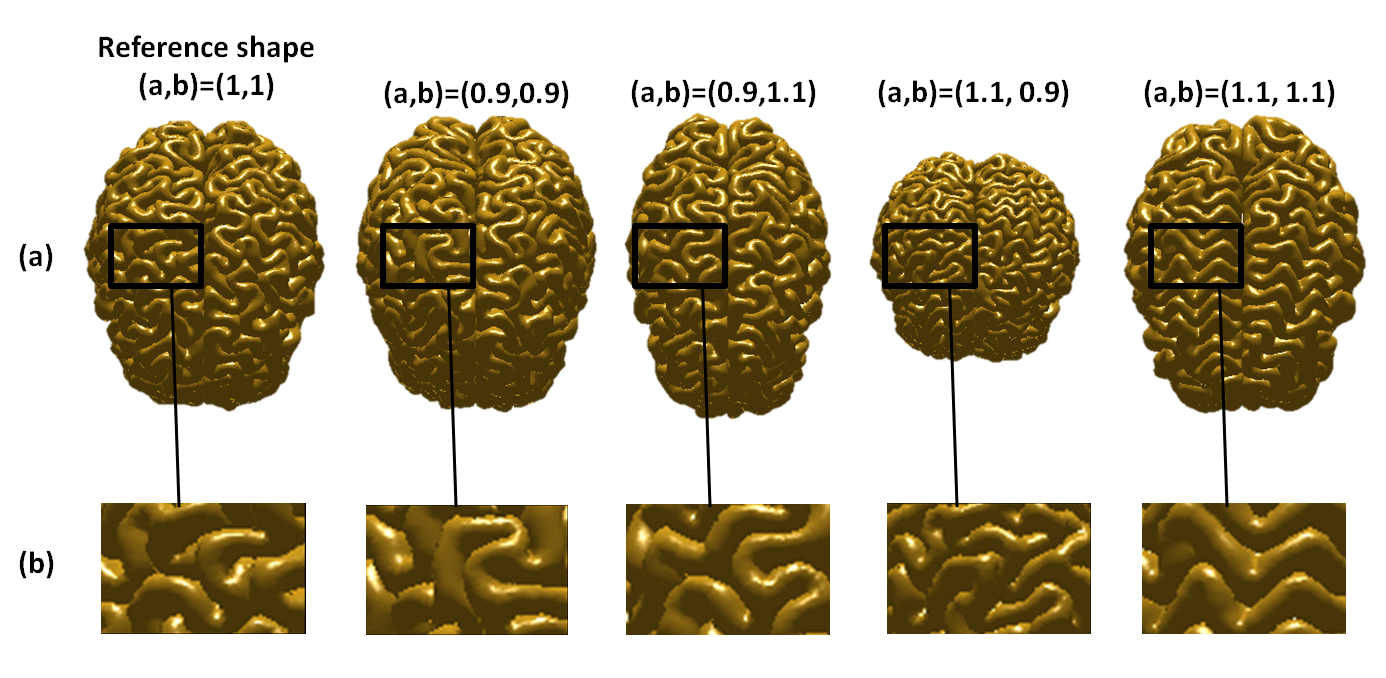}
\vspace{-0.5cm}
\caption{Variations in shape, size, placement and orientation of cortical folds across simulations. (a) Simulated fetal brains at step 22000 with $(a,b)=(0.9,1,1.1)$. (b) Zoom of the part of cortical pattern inside the black rectangle.}
\label{fig:folds}       
\end{figure}

These results show that the variations in the initial geometry of the brain strongly influences cortical folding patterns, either in terms of shape, size, placement and orientation of cortical folds. Zoomed examples are illustrated in Fig. \ref{fig:folds}.b.

\section{DISCUSSION}
\label{sec:discuss}

In this paper, we have proposed a framework for comparing a large number of biomechanical simulations of human brains using an adaptive spherical parameterization approach, in order to study the impact of the initial geometry of the human brain on cortical patterns. The major advantage of our framework is that it enables to exploit both, realistic mechanical properties of the human brain and the global characteristics of the cortical surface via shape descriptors based on a spherical mapping. A future utility of our framework is that it can also allow us to compare these simulated cortical surfaces with real ones, and consequently to measure efficiency of a biomechanical model in terms of generating folds at the right positions and having forms consistent with that of real ones. A more detailed study of these cortical patterns will be presented in a forthcoming paper, in which we will also analyze the folds orientations changes using an estimation of the principal curvatures directions. 

\addtolength{\textheight}{-12cm}   



\bibliographystyle{./IEEEtran}
\bibliography{IEEEabrv,IEEEexample}

\end{document}